\documentstyle{article}
\begin{document}

\title{Approach to the Singularity in Spatially Inhomogeneous
Cosmologies} 
\author{Beverly K. Berger \\
Department of Physics, Oakland University, Rochester, MI 48309 USA}

\maketitle
\bigskip
\begin{abstract}
A combination of analytic and numerical methods has yielded a clear understanding of the
approach to the singularity in spatially inhomogeneous cosmologies. Strong support is found for the
longstanding claim by Belinskii, Khalatnikov, and Lifshitz that the collapse is dominated by local
Kasner or Mixmaster behavior. The Method of Consistent Potentials is used to establish the
consistency of asymptotic velocity term dominance (AVTD) (local Kasner behavior) in that no terms
in Einstein's equations will grow exponentially when the VTD solution, obtained by neglecting all
terms containing spatial derivatives, is substituted into the full equations. When the VTD
solution is inconsistent, the exponential terms act dynamically as potentials either to drive the
system into a consistent AVTD regime or to maintain an oscillatory approach to the singularity.
\end{abstract}

\section{Introduction}
This Chapter, written from the physicist's point of view, will describe an example of
synergy between mathematics, physics, and numerical simulation. The work reported here was done in
collaboration with David Garfinkle (Oakland University), James Isenberg (University of
Oregon), Vincent Moncrief (Yale University), and Marsha Weaver (Max Planck Institute for
Gravitational Physics) \cite{berger98c}. Our objective was to explore the approach to the
singularity in spatially inhomogeneous cosmologies.

Much earlier (in the 1960's), Belinskii, Khalatnikov, and Lifshitz (BKL)
\cite{belinskii69a,belinskii69b,belinskii71a,belinskii71b,belinskii82} had provided a heuristic
analysis of such models suggesting that each spatial point would evolve toward the singularity as
if it were a spatially {\it homogeneous} cosmology. We shall call this behavior ``locality.'' 
In addition to the BKL claim of locality, there was also their claim that generically the
local approach to the singularity was that of the homogeneous Bianchi Type IX (or
Mixmaster) model \cite{misner69,ryan75}. The BKL claims were questioned and it was never
clear whether or not they were correct \cite{barrow79}.

To understand these remarks, we begin with the simplest cosmology --- the spatially
homogeneous, isotropic (all points and directions in {\it space} are equivalent)
Friedmann-Robertson-Walker models believed to describe our Universe on the
largest scales \cite{misner73,wald84}. Such models are characterized (if one considers evolving
backward in time) by infinite density, temperature, and tidal gravitational force a
finite time ago. This is the Big Bang singularity and means that if $t_0$ is the present
value of comoving proper time labeling a spacelike hypersurface $\Sigma_0$, there exists
some $\Sigma_i$ labeled by $t = 0$ such that $\Sigma_i$ marks a boundary of the spacetime.

More general, but still spatially homogeneous, cosmologies may be found by replacing the
isotropic universe expansion rate $H = \dot R / R$ by anisotropic rates $\{ H_x, H_y, H_z
\}$. The simplest of these is the Kasner or Bianchi Type I model where $t \sum_{i=1}^3
H_i = 1 = t^2 \sum_{i=1}^3 H_i^2$ \cite{kasner25}. These models are called velocity term
dominated (VTD) because the only dynamically relevant terms contain time derivatives.
Spatial derivatives can play a role in spatially homogneous cosmologies in the
computation of the spatial curvature in a coordinate frame. In Bianchi Type I models, the
spatial scalar curvature, ${}^3\!R$, is zero. However, it is non-zero for the other
Bianchi Types. The evolution of these models may be understood as dynamics in a
configuration space where each point is a cosmology at an instant of time. This space
will be finite dimensional with directions corresponding to each free function needed to
describe the space. The space of all 3-geometries (3-metrics modulo diffeomorphisms) is
called superspace \cite{wheeler68}. The restriction to homogeneous cosmologies is called
minisuperspace (MSS) \cite{misner72}. 

The spatial scalar curvature ${}^3\!R$ (actually ${}^3\!g \, {}^3\!R$ where ${}^3\!g$ is the
determinant of the spatial metric) acts as a potential for the dynamics in MSS. As one
approaches the singularity, any cosmology in this class behaves as a ``free particle''
--- i.e.~Kasner model --- until a ``bounce'' off the potential occurs. After the bounce,
the model behaves as a different Kasner universe. Homogeneous cosmologies based on most
of the Bianchi Types (in the vacuum case) are asymptotically VTD (AVTD) --- there is a
final MSS bounce. A fixed set of asymptotic Kasner parameters describes the approach to
the singularity. We note here that topological distinctions among the various Bianchi
spaces are dynamically irrelevant. Bianchi Types VIII and IX are not AVTD. The spatial
scalar curvature forms a closed MSS potential so that there is no last bounce. 
(Interestingly, while this statement is strongly supported by heuristic and numerical
evidence, there is no rigorous proof. The strongest results so far are those of Rendall
\cite{rendall97b} and Weaver \cite{weaver99b}.) The relationship between one Kasner and the next in
subsequent bounces can be expressed as a 4-parameter map which displays sensitivity to
initial conditions \cite{belinskii71b,chernoff83,berger96c}.  Such models are called Mixmaster
universes (after a food processor)
\cite{misner69}.

The BKL claim, then, that the approach to the singularity exhibits local Mixmaster
dynamics (LMD) means the following:  if, in some suitable but unspecified spacetime
foliation, one could write down the general solution to the vacuum Einstein equations,
the solution at any spatial point as a function of time would be arbitrarily close to a
Mixmaster universe. The solution looks like a Mixmaster universe with spatially dependent
Kasner parameters. For this to be true, at the very least, the time evolution of the
spatial dependence must be dynamically negligible compared to the local evolution. The
main criticism of the BKL claim has been that a specific spacetime slicing is imposed and
that global issues are not addressed. 

In the work discussed here, a combined analytic and numerical approach provides strong
support for the BKL claim. The analytic aspect of our methods, the Method of Consistent
Potentials (MCP) \cite{grubisic93}, is a ``consistency check'' for an AVTD approach to the
singularity. With a suitable choice of variables in some spacetime slicing, one may obtain the VTD
solution to Einstein's equations by solving the system of ODE's which remains if all spatial
derivatives are neglected. In some limit (say $\tau \to \infty$), the chosen variables
are linear or constant in $\tau$ (presumably with spatially dependent coefficients). The
VTD solution is then substituted into all the previously neglected terms. If these are
exponentially small as $\tau \to \infty$, then it is consistent that the model is AVTD.
Otherwise, one expects that the approach to the singularity is not AVTD. Surprisingly,
perhaps, the MCP predictions are borne out by detailed numerical simulations of the full
Einstein equations for these models
\cite{berger98c,berger93,berger97b,berger97e,berger98a,weaver98}.

In recent years, it has become possible to prove that the behavior of certain simple
spatially inhomogeneous cosmologies is AVTD. An example is the polarized Gowdy model
consisting of a single polarization of gravitational waves with symmetry in the plane
perpendicular to the direction of propagation. The wave amplitude satisfies a linear
equation while the rest of the spacetime may be constructed if the wave amplitude is
known. The addition of the second polarization completely changes the character of the
solution. The two wave amplitudes satisfy nonlinearly coupled wave equations. Numerical
solutions, as interpreted through the MCP, for thses models have allowed sufficient
understanding of the solutions so that rigorous study is possible. It can now be
proved that solutions with the numerically observed behavior exist \cite{kichenassamy98}. This is
not a redundant effort because the proof overcomes objections concerning numerical artifacts or
limitations on the ability to explore initial data space. Furthermore, the numerical
studies cannot be carried to $\tau = \infty$.

In Section II, we shall present an introduction to homogeneous cosmologies including a
detailed example of the use of the MCP. In Section II, we shall discuss Gowdy models ---
showing how some are AVTD, interesting features of the approach to the singularity, and
how adding a magnetic field yields LMD \cite{berger93,berger97b,weaver98}. In Section IV, we
discuss progress to date on the most complicated cosmologies we have studied --- those with a
single spatial $U(1)$ symmetry (which I propose to call Moncrief models)
\cite{moncrief86,berger97e,berger98a} ---and show the relation to the homogeneous Mixmaster model.
Discussion will be given in Section V.

\section{Spatially homogeneous cosmologies}
Bianchi's classification of homogeneous 3-spaces \cite{bianchi97} implies that they can be
described by metric 1-forms, $\sigma^i$, such that (for $i,j = 1,2,3$)
\begin{equation}
\label{bianchitype}
d \sigma^i = C^i_{jk} \sigma^j \wedge \sigma^k
\end{equation}
where $C^i_{jk}$ are the structure constants of a group (e.g. $SU(2)$ for Bianchi Type IX). A
homogeneous spacetime then has the metric
\begin{equation}
\label{bianchimetric}
ds^2 = -N^2 d\tau ^2 + e^{2\Omega} \left( e^{2 \beta} \right)_{ij} \sigma^i \sigma^j
\end{equation}
where
\begin{equation}
\label{betadef}
\beta_{ij} = {\rm diag} \left( -2 \beta_+, \beta_+ + \sqrt{3} \beta_-, \beta_+ - \sqrt{3}
\beta_- \right) \ .
\end{equation}
The lapse $N(\tau)$ allows a change of time coordinate. The spacetime foliation is chosen to make
the homogeneity manifest. We shall follow BKL and take $N = e^{3 \Omega}$. The chosen variables,
$\Omega(\tau)$ and $\beta_\pm(\tau)$ respectively, measure the volume and anisotropy (two
orthogonal shears) of the space at a given time. MSS is the 3-dimensional space with axes $(\Omega,
\beta_+,\beta_-)$. Each point in MSS is a homogeneous space. A trajectory in MSS is a
cosmological spacetime.

To understand the dynamics in MSS, it is most convenient to use the Hamiltonian
formulation of Einstein's equations pioneered by Arnowitt, Deser, and Misner (ADM)
\cite{arnowitt62,misner73,ryan75}. When specialized to homogeneous spaces, Einstein's equations for
Class A Bianchi types (the only ones we shall consider here) may be obtained by variation of the
``superhamiltonian,'' $NH$, where $H = 0$ is the Hamiltonian constraint. For the Kasner
and Bianchi Type IX Mixmaster models we find
\begin{equation}
\label{mixh0}
2 H = - p_\Omega^2 + p_+^2 + p_-^2 + V(\beta_\pm, \Omega)
\end{equation}
where
\begin{equation}
\label{vmss}
V=\left\{ \matrix{0\quad \quad \quad , \quad \quad {\rm Bianchi\  Type\  I}\hfill\cr
V_{IX} \quad \quad , \quad \quad {\rm Bianchi\  Type\  IX} \hfill\cr} \right.\ 
\end{equation}
for
\begin{eqnarray}
\label{vix}
V_{\rm IX} &=&  e^{4\Omega }[e^{-8\beta _+}+e^{4(\beta _++\sqrt 3\beta _-)}+
e^{4(\beta _+-\sqrt 3\beta _-)} \nonumber \\
& &-2e^{4\beta _+}-2e^{-2(\beta _++\sqrt 3\beta
_-)}-2e^{-2(\beta _+-\sqrt 3\beta _-)}]
\end{eqnarray}
while $(p_\Omega,
p_\pm)$ are the momenta canonically conjugate to $(\Omega, \beta_\pm)$. The VTD solution
is just the Kasner solution
\begin{eqnarray}
\label{kasner}
\Omega &=& \Omega_0 - p_\Omega \tau \ , \nonumber \\
\beta_\pm &=& \beta_\pm^0 + p_\pm \tau
\end{eqnarray}
with (from $H=0$)
\begin{equation}
\label{kasnerh0}
p_\Omega^2 = p_+^2 + p_-^2.
\end{equation}
To apply the MCP to the Bianchi Type IX model, substitute the Kasner solution
(\ref{kasner})-(\ref{kasnerh0}) into (\ref{vix}) to give (as $\tau \to \infty$)
\begin{eqnarray}
\label{vixlim}
V_{IX}^{\rm lim} &=& e^{4 |p_\Omega| \tau (-1 + 2 \cos \theta)}+ e^{4 |p_\Omega| \tau (-1 -
 \cos \theta + \sqrt{3} \sin \theta)} \nonumber \\
 & &+ e^{4 |p_\Omega| \tau (-1 -
 \cos \theta - \sqrt{3} \sin \theta)}  + \ldots
\end{eqnarray}
where $\cos \theta = - p_+ / p_\Omega$ and $\sin \theta = - p_- / p_\Omega$. The terms
indicated by $\ldots$  are almost always negligible \cite{barrow82,berger91} and we shall ignore
them here. Clearly, except for the set of measure zero $\theta = (0, \pm 2\pi/3)$, one of the terms
on the right hand side of (\ref{vixlim}) will grow. This leads to the MCP prediction that
generic Bianchi Type IX models are not AVTD. This, of course, is consistent with the
belief that there is no last bounce in these models.

Asymptotically, the Bianchi Type IX model solution may be described as a presumably
infinite sequence of Kasner parameters. It was shown long ago by BKL \cite{belinskii71b} that the
Kasner exponents $\{p_i\}$, such that the anisotropic scale factor $H_i = t^{p_i}$, may be
expressed in terms of a single parameter
$1 \le u < \infty$ via
\begin{eqnarray}
\label{kasnerindices}
p_1 & = & -u / (u^2 + u + 1) \ , \nonumber \\
p_2 & = &(u+1) / (u^2 + u + 1) \ , \nonumber \\
p_3 & = & u(u+1) / (u^2 + u + 1) 
\end{eqnarray}
where
\begin{equation}
\label{umap}
u_{n+1}=\left\{ \matrix{u_n-1\quad \quad ,\quad \ \  u_n\ge 2\hfill\cr
  (u_n-1)^{-1}\;\;\,,\quad 1\le u_n\le 2\hfill\cr} \right.
\end{equation}
for $u_n$ the $u$-parameter of the $n$th Kasner epoch. This map contains the information
in the Gauss map
\begin{equation}
\label{gaussmap}
u_{N+1} = {1 \over {u_N - [ u_N]}}
\end{equation}
where $[\ ]$ denotes integer part. The Gauss map is known to be chaotic \cite{barrow82}. To
arbitrary accuracy, as long as the terms indicated by \dots in (\ref{vixlim}) are negligible, an
approximate evolution may be constructed using the 4-parameter map
\cite{belinskii71b,chernoff83,berger96c} as a sequence of Kasner models with the rules to go from
one Kasner model to the next. From this analysis, one can approximate the duration of each Kasner
epoch (in $\tau$ or $|\Omega|$). This quantity depends on the values of the parameters but
increases roughly exponentially in $\tau$ toward the singularity. 

If the BKL claim is correct, asymptotically each spatial point would be characterized by
a spatially dependent $u(x)$, obeying (\ref{umap}) to ever better accuracy in the
approach to the singularity. But the Kasner epoch duration dependence on $u(x)$ means
that bounces occur at different times at different spatial points, leading eventually to
a complicated and arguably fractal spatial dependence of the metric \cite{belinskii92,kirillov87}.

For later reference, we note here that the addition of magnetic fields to
AVTD Bianchi Type vacuum cosmologies can ``close'' the MSS potential to yield a Mixmaster
model \cite{jantzen86}. One example is the magnetic Bianchi VI$_0$ cosmology with
\cite{leblanc95,berger96a}
\begin{eqnarray}
\label{hvi0}
2H_{{\rm VI}_0} &=& -p_\Omega^2 + p_+^2 + p_-^2 + B^2 e^{2 \Omega - 4 \beta_+} +e^{4 \Omega +
4 \beta_++4 \sqrt{3} \beta_-} \nonumber \\
& &+e^{4 \Omega + 4
\beta_+-4 \sqrt{3} \beta_-}+ \ldots
\end{eqnarray}
where \dots indicates terms different from those in (\ref{vixlim}) but still dynamically
irrelevant almost always.

\section{Gowdy cosmologies and their generalizations}
The simplest solutions representing spatially inhomogeneous cosmological spacetimes are
the polarized Gowdy models on $T^3 \times R$ \cite{gowdy71,berger74} described by the metric
\begin{equation}
\label{gowdypolmetric}
ds^2 = e^{\lambda/2}(-e^{-3\tau/2}\,d\tau^2 + e^{\tau/2} \, d\theta^2) + e^{-\tau} (e^{P}
d\sigma^2 + e^{-P} \, d\delta^2) 
\end{equation}
where $\lambda$ and $P$ are functions of $0 \le \theta \le 2 \pi$ and time, $\tau$, only.
Here $P$ satisfies a linear wave equation
\begin{equation}
\label{peqpol}
P,_{\tau \tau} - e^{-2 \tau} \, P,_{\theta \theta} = 0
\end{equation}
with explicit solution
\begin{equation}
\label{psolnpol}
P = P_0 + \alpha_0 \tau + \sum_{n=1}^\infty \, \alpha_n {\cal Z}_0(ne^{-\tau})
\cos(n\theta + \phi_n)
\end{equation}
where ${\cal Z}_0(x)$ is a zero-order Bessel function. Given $P$, $\lambda$ may be
constructed from the Hamiltonian and momentum constraints written respectively as
\begin{equation}
\label{h0pol}
\lambda,_\tau = -(P,_\tau)^2 - e^{-2 \tau} (P,_\theta)^2,
\end{equation}
\begin{equation}
\label{hqpol}
\lambda,_\theta = - 2 P,_\tau P,_\theta.
\end{equation}
This model may be obtained as an inhomogeneous generalization of a Kasner solution. The
approach to the singularity was shown heuristically early on \cite{berger74} and later
rigorously \cite{isenberg90} to be AVTD. It was also proven that the solution may be extended to
the strong curvature singularity at $\tau = \infty$ and only becomes singular in that
limit \cite{chrusciel90}.

The Gowdy models become more interesting when the second gravitational wave polarization
is added \cite{moncrief81,berger93}. The metric becomes
\begin{eqnarray}
\label{gowdymetric}
ds^2 &=& e^{\lambda/2}(-e^{-3\tau/2}\,d\tau^2 + e^{\tau/2} \, d\theta^2) \nonumber \\
& &+ e^{-\tau}
[e^{P} d\sigma^2 + 2 Q e^{P} d\sigma d \delta + (Q^2 \, e^{P} + e^{-P}) \, d\delta^2] 
\end{eqnarray}
where $P$ and $Q$ satisfy the nonlinearly coupled equations
\begin{equation}
\label{peqgowdy}
P,_{\tau \tau} - e^{-2 \tau} P,_{\theta \theta} - e^{2P}(Q,_\tau^2 - e^{-2\tau}
Q,_\theta^2) = 0,
\end{equation}
\begin{equation}
\label{qeqgowdy}
Q,_{\tau \tau} - e^{-2 \tau} Q,_{\theta \theta} + 2 P,_\tau Q,_\tau - 2 e^{-2 \tau}
P,_\theta Q,_\theta = 0
\end{equation}
while the constraints become
\begin{equation}
\label{h0unpol}
\lambda,_\tau = -P,_\tau^2 - e^{2P} Q,_\tau^2 - e^{-2 \tau} (P,_\theta^2 + e^{2P}
Q,_\theta^2),
\end{equation}
\begin{equation}
\label{hqunpol}
\lambda,_\theta = - 2 (P,_\tau P,_\theta + e^{2P} Q,_\tau Q,_\theta).
\end{equation}
This model is still an inhomogeneous generalization of the Kasner solution but with a
rotation of the principle axes of the 3-space between one spatial point and another.

To apply the MCP, it is convenient to use the Hamiltonian density
\begin{equation}
\label{gowdywaveh}
2H=
\, \pi _P^2+\kern 1pt\,e^{-2P}\pi _Q^2 
  + \, e^{-
2\tau }\left( P,_\theta ^2+\, e^{2P}Q,_\theta ^2 \right) 
\end{equation}
whose variation yields the wave equations (\ref{peqgowdy}) and (\ref{qeqgowdy}). The first step is
to solve (\ref{peqgowdy})-(\ref{hqunpol}) neglecting all terms containing spatial derivatives. In
the limit as
$\tau
\to \infty$, we find the VTD solution \cite{berger93,berger97b}
\begin{equation}
\label{gowdyvtd}
P = v \tau \quad , \quad Q = Q_0 \quad , \quad \lambda = - v^2 \tau
\end{equation}
where $v$ depends on $\theta$ but not on $\tau$. Now substitute (\ref{gowdyvtd}) into
(\ref{gowdywaveh}). We see that
\begin{equation}
\label{gowdyv1}
V_1 = {1 \over 2} \, \pi_Q^2 \,  e^{-2P}
\end{equation}
can be exponentially small if $v > 0$ while
\begin{equation}
\label{gowdyv2}
V_2 = {1 \over 2} \, Q,_\theta^2 \, e^{2P-2\tau}
\end{equation}
can be exponentially small if $v < 1$. Thus the MCP predicts that the approach to the
singularity in these models will be AVTD (almost everywhere \cite{berger97b}) if $0 < v < 1$. 

The question then arises as to what happens if $v < 0$ or $v > 1$ in the initial data.
Prior to the asymptotic regime, spatial derivatives need not be dynamically irrelevant
and a variety of nonlinear effects may occur \cite{berger97b}. At some value of $\tau$ at a given
$\theta = \theta_i$, the true Hamiltonian is excellently approximated by either
\begin{equation}
\label{gowdyh1lim}
H_1 = {1 \over 2} \pi_P^2 + {1 \over 2} \pi_Q^2 e^{-2P}
\end{equation}
or
\begin{equation}
\label{gowdyh2lim}
H_2 = {1 \over 2} \pi_P^2 + {1 \over 2} Q,_\theta^2 e^{2P-2\tau}.
\end{equation}
In the first case, there will be a bounce off $V_1$ causing $\pi_P$ (and thus
$v(\theta_i)$) to change sign. But $V_1$ will cause a bounce only when it is
exponentially growing --- i.e.~when $v(\theta_i)$ is initially $<0$. The bounce will then
yield the new value $v'(\theta_i) > 0$. In the second case, $v(\theta_i)-1$ changes sign at the
bounce off $V_2$. Here the bounce occurs only for $v(\theta_i) > 1$ so that $V_2$ grows
exponentially. If $v'(\theta_i) - 1 = -(v(\theta_i)-1)$, then $v'(\theta_i) = 2 -
v(\theta_i)$. If $v'(\theta_i) < 0$, there will be another bounce off $V_1$ and so on. No
additional bounces will occur once $0 < v(\theta_i) < 1$. Thus the MCP predicts a
sequence of bounces off $V_1$ and $V_2$ at spatial points where $v(\theta) < 0$ or
$v(\theta) > 1$ until $0 < v(\theta) < 1$ at which point the true solution approaches a
VTD solution at these points.

There is a set of measure zero at which the coefficient of $V_1$ (i.e.~$\pi_Q$) or $V_2$
(i.e.~$Q,_\theta$) vanishes. At these spatial points, the above mechanism fails to
operate so that $v(\theta) > 1$ or $v(\theta) < 0$ is allowed \cite{berger97b}. If, at $\theta_0$,
$Q,_\theta = 0$, then, for $\theta \approx \theta_0$, $Q,_\theta \approx 0$ so that $v(\theta) > 1$
can persist for a long time. This creates ``spiky'' features in $P$ in the vicinity of
one of these non-generic points. Similarly, $P,_\tau < 0$ can persist if $\pi_Q \approx
0$. This in turn leads to an apparent ``discontinuity'' in $Q$ since $Q,_\tau = \pi_Q \,
e^{-2P}$ grows exponentially in the positive or negative direction about the region where
$\pi_Q$ goes through zero with $P,_\tau < 0$. 

All the properties described here in terms of the MCP have been observed in numerical
simulations of the full Einstein equations (\ref{peqgowdy})-(\ref{hqunpol})
\cite{berger93,berger97b}. This provides strong support for the BKL claim that each spatial point
evolves toward the singularity as a separate universe. So far, these models are too restrictive to
admit LMD. Once the behavior of these models became apparent in the numerical simulations,
Kichenassamy and Rendall \cite{kichenassamy98} were able to apply the former's methods to prove
that solutions with this property exist. Although the numerical simulations provide information
about the nature of generic solutions, they cannot explore all initial data space nor can they
evolve to $\tau = \infty$. The MCP provides a framework to interpret observed waveforms
and thus to identify generic behavior independent of initial conditions. Similarly, it
allows one to argue that nothing qualitatively new will happen if the simulations are
followed longer. While the studies in \cite{kichenassamy98} do not prove the solutions are AVTD
(in contrast to the result \cite{isenberg90} for polarized Gowdy models), they do provide some
rigorous verification of the numerical results.

To overcome the limitations of the vacuum Gowdy models, consider the generalization of
the metric (\ref{gowdymetric}) to represent the inhomogeneous version of the magnetic Bianchi
VI$_0$ model \cite{weaver98,weaver99a}. It is necessary to change the spatial topology from $T^3$
to the solv-twisted torus \cite{fujiwara93} with boundary conditions \cite{weaver98,weaver99a}
$\lambda (\theta + 2
\pi,
\tau) = \lambda(\theta, \tau)$,  $P (\theta + 2 \pi,
\tau) = P(\theta, \tau) + 2 \pi a$, $Q (\theta + 2 \pi,
\tau) = Q(\theta, \tau)\, e^{-2\pi a}$ where $a$ is a constant. To incorporate the
coupling to the magnetic field, the metric must be generalized and becomes
\begin{eqnarray}
\label{maggowdymetric}
ds^2 &=& e^{\lambda/2}(-e^{-3\tau/2}\,d\tau^2 + e^{(\mu +\tau)/2} \, d\theta^2) 
\nonumber \\
& &+ e^{-\tau}
[e^{P} d\sigma^2 + 2 Q e^{P} d\sigma d \delta + (Q^2 \, e^{P} + e^{-P}) \, d\delta^2] 
\end{eqnarray}
which yields the Hamiltonian density
\begin{eqnarray}
\label{hmaggowdy}
H &=& {1 \over {4 \pi_\lambda}} \left( \pi_P^2 + \pi_Q^2 \, e^{-2P} + e^{-2\tau}
P,_\theta^2 + e^{2(P-\tau)} Q,_\theta^2 \right) \nonumber \\
& &+ 4 \pi_\lambda e^{(\lambda +\tau)/2} \, B^2
\end{eqnarray}
where $B$ is a constant related to the magnitude of the magnetic field. The constraint
$\pi_\lambda = {\textstyle {1 \over 2}} e^{\mu/4}$ has been used to obtain this
expression \cite{weaver98,weaver99a}. The VTD solution is the same as (\ref{gowdyvtd}) for the
vacuum case if we make the substitution
\begin{equation}
\label{magneticpdot}
v(\theta) \equiv P,_\tau = {{\pi_P} \over {2 \pi_\lambda}}
\end{equation}
where $\pi_\lambda$ also depends on $\theta$ but not on $\tau$. The new definitions of
$V_1$ and $V_2$ require only division by $2 \pi_\lambda$. All the arguments regarding
$V_1$ and $V_2$ go through as before. However, if $0 < v(\theta) < 1$, the new potential
in (\ref{hmaggowdy}),
\begin{equation}
\label{v3mag}
V_3 = 4 \pi_\lambda e^{(\lambda + \tau)/2} \, B^2 \, ,
\end{equation}
will grow exponentially. Thus the dynamics will be dominated by one of $H_1$ (from
(\ref{gowdyh1lim})), $H_2$ (from (\ref{gowdyh2lim})) or
\begin{equation}
\label{hmag}
H_3 = {{\pi_P^2} \over {4 \pi_\lambda}} + 4 \pi_\lambda e^{(\lambda + \tau)/2} \, B^2 \,.
\end{equation}
Since no value of $v(\theta)$ is consistent with $V_1$, $V_2$, {\it and} $V_3$ all
exponentially small, the MCP predicts that the magnetic Gowdy model is not AVTD. Since
this magnetic Gowdy model was obtained from magnetic Bianchi Type VI$_0$, the observed
departure from AVTD behavior should indicate LMD. Numerical simulations support the MCP
predictions to the extent that they can be followed \cite{weaver98}. So far, nothing rigorous is
known about the solutions.

\section{$U(1)$ symmetric cosmologies (Moncrief models)}
Some years ago, Moncrief used an ADM-like approach to write Einstein's equations for a
cosmology with a single spatial $U(1)$ symmetry \cite{moncrief86}. He considered Einstein's
equations projected along and normal to the Killing direction. It is convenient to use a metric of
the form \cite{moncrief86}
\begin{eqnarray}
\label{u1metric}
ds^2&=&e^{-2\varphi }\left\{ -\tilde N d\tau ^2+
\tilde g_{ab}(dx^a + \tilde N^a d\tau ) (dx^b+ \tilde N^b d\tau ) \right\} \nonumber
\\
& &+e^{2\varphi } \left( dx^3+\beta _adx^a+\beta_0 d\tau \right)^2
\end{eqnarray}
where $x^3$ is the Killing direction, $x^1 = u$ and $x^2 = v$ are the other spatial
coordinates, while $\tau$ is the time. The variables are $\varphi$, where $e^\varphi$ is
the norm of the Killing field and $\beta^a$ for $a = 1,\,2$ are the ``twists.'' The
conformal lapse $\tilde N$ (chosen to be $\tilde N = e^\Lambda$), shifts $\tilde N^a$ (chosen to
vanish), and $\beta_0$ become Lagrange multipliers of the constraints. The spatial $U(1)$ symmetry
allows the last constraint,
$e^a,_a = 0$, to be satisfied identically through the introduction of the twist potential $\omega$
such that
$e^a = \epsilon^{ab} \omega,_b$ where $\epsilon^{ab} = -\epsilon^{ba}$. Thel 2-metric in
the $u$-$v$ plane is parametrized as
\begin{equation}
\label{2metric}
\tilde g_{ab} = e^\Lambda e_{ab}
\end{equation}
with
\begin{equation}
\label{eab}
e_{ab}={\textstyle{{1 \over 2}}}\left(
{\matrix{{e^{2z}+e^{-2z}(1+x)^2}&{e^{2z}+e^{-2z}(x^2-1)}\cr
{e^{2z}+e^{-2z}(x^2-1)}&{e^{2z}+e^{-2z}(1-x)^2}\cr }} \right) \ .
\end{equation}
If the variables $\{\varphi, \omega, \Lambda, x, z\}$ are joined by their canonically
conjugate momenta $\{p, r, p_\Lambda, p_x, p_z \}$, it is found that Einstein's equations
may be obtained by variation of the Hamiltonian density \cite{berger97e,berger98a}
\begin{eqnarray}
\label{Hu1}
H &=& {{{1 \over 8}}}p_z^2+{{{1 \over
2}}} e^{4z}p_x^2+{{{1 \over 8}}}p^2+{\textstyle{{1 \over 2}}}e^{4\varphi
}r^2-{{{1 \over 2}}}p_\Lambda ^2   \nonumber \\
&& + \left( {e^\Lambda e^{ab}} \right) ,_{ab}- \left( {e^\Lambda e^{ab}}
\right) ,_a\Lambda ,_b+e^\Lambda   \left[  \left( {e^{-2z}}
\right) ,_u x,_v- \left( {e^{-2z}} \right) ,_v x,_u \right] \nonumber \\
&& +2e^\Lambda e^{ab}\varphi ,_a\varphi ,_b+{1 \over 2}
e^\Lambda e^{-4\varphi }e^{ab}\omega ,_a\omega ,_b   
\end{eqnarray}
where $H = 0$ is the Hamiltonian constraint.
Since the structure of this Hamiltonian is reminiscent of that for the Gowdy models
(\ref{gowdywaveh}) and (\ref{hmaggowdy}), one may consider application of the MCP. The VTD solution
is
\begin{eqnarray}
\label{u1avtd} 
z&=&-v_z\tau,  \quad x= x_0,  \quad p_z= -4v_z, \quad
p_x= p_x^0, \quad
\varphi= -v_\varphi\tau, \nonumber  \\
\omega&=& \omega_0, \quad
p=-4v_\varphi,  \quad
r= r^0, \quad
\Lambda = \Lambda_0  - v_\Lambda\tau, \quad
p_\Lambda = v_\Lambda
\end{eqnarray}
where $v_z$, $v_\varphi$, $x_0$, $p_x^0$, $\omega_0$, $r^0$, $\Lambda_0$, and
$v_\Lambda > 0$ are functions of $u$ and $v$ but independent of $\tau$. (The sign of  
$v_\Lambda$ is fixed to ensure collapse.) 
Although quite complicated, the nonlinear terms in (\ref{Hu1}) fall into four classes:

(1) First, the term
\begin{equation}
\label{vzu1}
V_z = {1 \over 2} p_x^2 e^{4 z}
\end{equation}
grows exponentially only if $v_z < 0$;

(2) Second, we find that
\begin{equation}
\label{v1u1}
V_1 = {1 \over 2} r^2 e^{4 \varphi}
\end{equation}
grows exponentially only if $v_\varphi < 0$.

If we then assume $v_z > 0$, $v_\varphi > 0$, the inverse conformal 2-metric $
e^{ab}$ is dominated by $e^{-2z}$ which grows exponentially. 

(3) It is then seen that all but one term in (\ref{Hu1}) containing spatial derivatives has the
factor
\begin{equation}
\label{factor}
F = e^{\Lambda - 2z}\ .
\end{equation} 

(4) The remaining exponential term is
\begin{equation}
\label{v2u1}
V_2 = {1 \over 2} e^{\Lambda + 4 \varphi} e^{ab} \omega,_a \omega,_b \approx e^{\Lambda -
2z + 4 \varphi} ( \nabla \omega)^2 \ .
\end{equation}
The VTD form of the Hamiltonian constraint is
\begin{equation}
\label{hvtdu1}
H_{\rm VTD} = -{1 \over 2} p_\Lambda^2 + {1 \over 8} p^2 + { 1 \over 8} p_z^2 = -{1 \over
2} v_\Lambda^2 + 2 v_\varphi^2 + 2 v_z^2 = 0 \ .
\end{equation}
For $v_\Lambda > 0$ to ensure collapse, $v_\varphi > 0$, and $v_z > 0$, (\ref{hvtdu1})
means that $e^\Lambda e^{ab } \approx e^{\Lambda - 2z}$ will always be exponentially
small since $v_\Lambda \ge 2 v_z$ while $V_2$ will always grow exponentially since
$v_\Lambda < 2 v_z + 4 v_\varphi$. This means that $v_\varphi > 0$, $v_z > 0$ is
inconsistent and a bounce off $V_2$ will always occur. The ``bounce rules'' ---
equivalent to conservation of momentum --- are \cite{berger99b}
\begin{eqnarray}
\label{bouncerules}
v'_\varphi &=& - v_\varphi + {1 \over 2} v_\Lambda - v_z \ , \nonumber \\
v'_z &=& - v_\varphi + {1 \over 4} v_\Lambda + {1 \over 2} v_z  \ ,\nonumber \\
v'_\Lambda &=& - 2v_\varphi + {1 \over 2} v_\Lambda - v_z 
\end{eqnarray}
indicating that $v_\varphi$ changes sign at the bounce. Occasionally, if $v_z >
2 v_\varphi - {1 \over 2} v_\Lambda $, $v_z$ can also change sign. If either $v_z$ or
$v_\varphi$ is $< 0$, a bounce off $V_z$ or $V_1$ will subsequently occur to reverse the sign.
Thus the MCP predicts an infinite sequence of bounces (almost) everywhere. (The behavior
near non-generic spatial points has not yet been studied for these models.)

The condition $r = \omega = 0$ is preserved by
Einstein's equations and yields polarized Moncrief models. In these models, both $V_1$
and $V_2$ are identically zero. If $v_z < 0$ initially at some spatial point, the MCP
predicts that a bounce off $V_z$ will occur to yield $z$ increasingly large and negative.
Since the surviving exponential terms with factor $F$ are exponentially small under these
circumstances, the MCP predicts that polarized Moncrief models should be AVTD. Numerical
simulations for these models do not have the problematic numerical aspects of generic Moncrief
models and yield strong evidence to support the MCP prediction \cite{berger97e,berger98a}. The
methods of Kichenassamy can be used to {\it prove} that these models are AVTD \cite{isenberg00}.

The constraints in the Moncrief models are non-trivial and must be solved
initially. In published work, we have used the following algebraic prescription
to solve the constraints \cite{berger97e}: Solve the momentum constraints (see
\cite{moncrief86,berger97e}) by imposing
$p_x = p_z = \varphi,_a = \omega,_a = 0$ and $p_\Lambda = c e^\Lambda$. For $c$
sufficiently large, the Hamiltonian constraint may be solved for $p$ or $r$. Thus
$x$, $z$, $\lambda$, and $r$ or $p$ may be freely specified. 

Numerical simulations of
these models are less straightforward than in the Gowdy case. In order to be able to
follow the simulations toward the singularity, it appears to be necessary to prevent the
development of Gowdy-like spiky features by introducing data smoothing
\cite{berger98a} --- every variable is replaced by a 6th order accurate average of itself over a
grid of nearby spatial points. In addition, the crucial and explicit role of $H_{\rm
VTD}$ in the dynamics means that it is essential to preserve the Hamiltonian constraint.
This is done by solving the Hamiltonian constraint, $H$ (from (\ref{Hu1})) $=0$, for
$p_\Lambda$ at every time step. We believe that both these numerical procedures not
only preserve a stable evolution and but also preserve the qualitative features ---
especially the evolution in time at a fixed point in space --- of the actual solution
\cite{berger98a}. 

Once again the numerical simulations support the MCP predictions. This is perhaps the
most remarkable case since, for the MCP to yield valid predictions, the time scale for
the change in the coefficients of the exponentials in $V_1$, $V_z$, and $V_2$ must be
significantly longer than that for the exponential itself.

Aside from the need for improved numerics, the remaining major open question is
whether the oscillations in $\varphi$ observed in Moncrief models correspond to LMD. This
issue may be addressed by writing a homogeneous Mixmaster model (on $S^3$) in the Moncrief
(on $S^3$) variables --- i.e.~identify the relation between the variables in the metric
(\ref{u1metric}) and those in the metric (\ref{bianchimetric}) by writing the metric 1-forms
$\sigma^i$ in a coordinate basis as $\sigma^1 = \cos \phi \, d\theta + \sin \theta \sin \phi \, d
\psi$, $\sigma^2 = \sin \phi \, d\theta - \sin \theta \cos \phi \, d\psi$, $\sigma^3 = d\phi +
\cos \theta d\psi$ \cite{misner73}. It is then straightforward to identify the cyclic coordinate
$\psi$ with the $U(1)$ symmetry direction $x^3$. Details of the transformation between Moncrief
and MSS variables are given elsewhere \cite{grubisic94,berger99b}.

For our purposes here, it is important to consider the Mixmaster models in terms of the
logarithmic scale factors
\begin{eqnarray}
\label{scalefactors}
\alpha &=& \Omega - 2 \beta_+ \, , \nonumber \\
\zeta &=& \Omega + \beta_+ + \sqrt{3} \beta_- \, , \nonumber \\
\gamma &=& \Omega + \beta_+ - \sqrt{3} \beta_- \, 
\end{eqnarray}
--- i.e.~ $e^\alpha$ is the 1-direction scale factor, etc. 
In a regime such that
$\alpha > \zeta > \gamma$, then
\begin{equation}
\label{phimix}
\varphi = \alpha + {1 \over 2} \ln \left[ \sin^2 \theta \sin^2 \phi + e^{2(\zeta -
\alpha)} \cos^2 \phi \sin^2 \theta + e^{2(\gamma - \alpha)} \cos^2 \theta \right] \, .
\end{equation}
This expression is exact but is written to express the dominance of the largest
logarithmic scale factor (LSF). The result for any other ordering of the scale factors is
obtained by the appropriate permutation. The identification of $\varphi$ with the largest
LSF creates an oscillatory behavior in $\varphi$ which may be understood from the behavior of
the LSF's in Mixmaster dynamics. For $\alpha > \zeta > \gamma$, $\alpha$ increases while $\zeta$
decreases until there is a bounce. At this bounce, $\dot \alpha$ and $\dot \zeta$ (where $\dot {}
\equiv d/d\tau$) change signs so that $\dot \alpha$ decreases and $\dot \zeta$ increases. The
Moncrief variable $\varphi$ follows $\alpha$ as long as $\alpha > \zeta$ --- through a sign change
of
$\varphi,_\tau$ at the bounce. When $\alpha = \zeta$, there is another sign change in
$\varphi,_\tau$ as it starts to track $\zeta$. While this is going on, $\gamma$
decreases monotonically. At the bounce, $\dot \gamma$ becomes less negative.
Eventually (when $1 < u < 2$ in (\ref{umap})), $\dot \gamma$ will change sign.

The Moncrief variable $z$ is written in terms of logarithmic scale factors as
\begin{equation}
\label{zmix}
z = {1 \over 2} (\gamma - \zeta) + {1 \over 2} \log \left[ {{\sin^2 \theta | \sin \phi| ( 1 +
\ldots) } \over {1 - \cos 2 \theta + \ldots}} \right]
\end{equation}
where terms indicated by $\ldots$ are exponentially small if $\alpha > \zeta > \gamma$. We see
that $z$ is large and negative unless $\gamma \approx
\zeta$ ---i.e.~$\dot \gamma > 0$ so that $\gamma$ is increasing. This marks the end of the
BKL era. Observation in numerical simulations of an increase in $z$ would provide a
qualitative signature for LMD which would be independent of numerical uncertainties. 

The usual Mixmaster bounces show up in the Moncrief models as bounces of $\varphi$
off $V_1$. The ``extra'' bounces in $\varphi$ are the bounces off $V_2$ which is not
exponentially small only when $\alpha \approx \zeta$ (for our original ordering of the
LSFs). The bounce in $z$ to halt its descent to large negative values also occurs off
$V_2$ and can be understood from the $V_2$ bounce laws (\ref{bouncerules}). The increase in $z$
will eventually be halted by a bounce off $V_z$.   

\section{Conclusions}
Despite the complexity of Einstein's equations, strong statements may be made about the
approach to the singularity. To make these statements, we have obtained a synergy among
heuristic, numerical, and mathematical analyses. The MCP has been applied to spatially
inhomogeneous cosmologies with as few as one spatial Killing vector field to predict whether
or not the approach to the singularity is AVTD. The observed numerical behavior has
already been used in the vacuum Gowdy and polarized Moncrief cases to guide the proof of
theorems.

The original BKL picture remained difficult to understand because BKL used more of the
VTD solution than was really necessary. Our consideration of the consistency of linear
asymptotic behavior in the MCP clarified the picture considerably. It is now possible to
determine quickly, given the appropriate choice of variables, how a collapsing model
should behave.

This brings us to the question of whether our results are dependent on the choice of
variables and the spacetime slicing. As far as the former goes, we now have two
completely different descriptions of the Mixmaster universe --- one in terms of the MSS
variables $\beta_\pm$, $\Omega$ and the other in terms of the Moncrief variables
$\varphi$, $z$, etc. In both cases, the VTD solutions for the variables are linear or
constant in $\tau$. (The same spacetime slicing is used in both descriptions.) In both
cases, there are exponential potentials which cannot be kept small. However, as we have
seen, the meaning of the variables and potentials is not the same in both cases. When
applied to the inhomogeneous models, we may conjecture that, even within a given
spacetime foliation, there may be many choices of variables which can yield an MCP
prediction and that the prediction --- AVTD or not --- will be independent of the choice.
Certainly, one should be able to construct analogs of the MSS variables for the Moncrief
models.

One might further conjecture that all reasonable foliations will converge asymptotically
in the sense that, e.g.~in a solution which is AVTD, in some foliation, it will be
possible to choose a time coordinate such that there are local Kasner variables linear in
that time coordinate. Some analytic work on polarized Moncrief models and Gowdy models
supports this conjecture.

We emphasize here the synergy between the MCP, the numerical simulations, and the
mathematics. To have only the simulations would suffer the limitations of restricted
initial data and limited duration. The MCP provides an organizing principle to allow
conjectures to be made about generic behavior based on the validity of the MCP as applied
to the simulations. Conjectures about the generic behavior then form the basis for
mathematical analysis.

So far, we have explored inhomogeneous cosmologies with one and two Killing fields. The
observed local behavior in all cases suggests that the boundary conditions are not
important and that the results are applicable in any collapse --- e.g.~in asymptotically
flat spacetimes. Some support for this exists in the observations of typical Gowdy
behavior in $S^2 \times S^1$ rather than $T^3$ spatial topology \cite{garfinkle99} and in the same
generic Mixmaster behavior in magnetic Gowdy models on the solv-twisted torus or $T^3$
\cite{weaver98,weaver99a,berger99a}.

One remaining open question in the Moncrief models is the extent to which the failure to
be AVTD indicates LMD. Writing a Bianchi Type IX Mixmaster cosmology as a Moncrief model provides a
qualitative signature for LMD in the rise of $z$. This may have been observed \cite{berger99b}. So
far, numerical difficulties make it impossible to conclude that every feature in the
asymptotic regime is LMD. Further studies are in progress.

What about the zero Killing field case? The local behavior may allow this problem to be
tractable numerically since high spatial resolution would not be required. There is no
obstruction to use of the MCP in this case. In fact, Moncrief has obtained the
Hamiltonian for this sytem starting from the metric (\ref{u1metric}) but assuming dependence on
$x^3$ as well as $u$, $v$, and $\tau$. The transformation to the twist potential $(r,
\omega)$ now cannot be made yielding the expected six degrees of freedom $\{\varphi, \beta_a,
x, z, \Lambda \}$.

Every term present in the Moncrief model is present in the more general case so that LMD
should be possible. It is not yet clear to what extent there are additional terms which
may signal a richer asymptotic behavior or whether the BKL picture will still hold in the
most general case.

\section*{Acknowledgments}
This work was supported in part by National Science Foundation
Grant PHY-9800103. 


\end{document}